%% file: main.tex
\documentclass{article}
\usepackage{spconf,amsmath,amssymb,graphicx,booktabs,array,cite,pifont}
\usepackage[table]{xcolor}
\usepackage{tikz}
\usetikzlibrary{arrows.meta,positioning}
\usepackage[hidelinks]{hyperref}
\usepackage{enumitem}
\usepackage{placeins}
\setlist{nosep, leftmargin=12pt}

\definecolor{ours}{RGB}{233,241,251}
\definecolor{tag}{RGB}{253,228,214}

\input{numbers}

\newcommand{\yes}{$\checkmark$}
\newcommand{\no}{$\times$}

\title{Measure Before You Repair: An Explainable Agent for Blind Restoration of Multi-Artifact 3-D MRI}
\name{Necati Sefercioglu$^{1,3}$, Mehmet Ozan Unal$^{2,3}$, Metin Ertas$^{3}$, Altan Cakir$^{1}$, Isa Yildirim$^{2,3}$}
\address{$^{1}$Istanbul Technical University, Dept.\ of Data Science and Analytics, Istanbul, Turkiye\\
$^{2}$Istanbul Technical University, Electronics and Communication Engineering Dept., Istanbul, Turkiye\\
$^{3}$Istanbul Technical University, Biomedical Imaging \& AI Lab, Istanbul, Turkiye}

\begin{document}
\ninept
\maketitle

\begin{abstract}
A brain MRI scan can carry an unknown mix of artifacts, so restoration has to be blind, yet most
deep-learning methods are built to correct one known artifact. We present Dx-Rx, an agent in which a multimodal large
language model (MLLM) measures suspected artifacts before it repairs them. To our knowledge, it is the first agent to
restore whole 3-D MRI volumes, whereas earlier MRI agents restore 2-D slices. In a loop, it looks at slices of the 3-D
volume, calls five diagnostic tools (noise, spike, ghosting, bias field, motion) and four restoration tools in any
order it chooses, checks each repair, can return to any earlier version to change the order, and ends with a
radiology-style report of its diagnostic reasons, the repairs it applied and their results. We simulated up to five
artifacts on \nTest{} whole \mbox{3-D} T1-weighted volumes, at levels calibrated to radiologists' grades of real
scans. An MLLM that only looked at the image, as earlier MRI agents do, removed at most \remImgHi{} of the artifact
error. With the best of five MLLMs, Dx-Rx removed a mean \remTerra, nearly matching the \remOracle{} of the same tools
given the ground-truth artifacts, which are never known for a clinical scan. With all five
models, it made no volume worse and was on par with or better than an expert rule-based pipeline built from the same
tools (\remRule). Code and demo are public at \url{https://itu-biai.github.io/measure-before-you-repair}.
\end{abstract}


\section{Introduction}
\label{sec:intro}
Artifacts from thermal noise, radio-frequency spikes, ghosting, coil shading (bias field) and head motion degrade
brain MRI and can force a rescan~\cite{zaitsev2015,narai2022}. A scan may carry several of them at once, and which ones are present
is not known in advance. The restoration has to be blind. Deep learning has been very successful at correcting one artifact, but
such methods do not generalize beyond it, which limits them in the clinic where several occur together~\cite{tang2025resmap}. Running every repair on every scan is not safe: on our test
set it made \harmApplyAll{} volumes worse. An expert pipeline instead runs each repair only when an artifact is detected,
in a fixed order. These rules are set by hand and stay fixed, so the pipeline cannot reorder its repairs when one interferes with another.

Agents built on large language models, which reason and call tools in a loop~\cite{yao2023react}, offer another
way. AgenticIR restores natural images with a language model that plans and reschedules and a vision-language model that judges quality~\cite{zhu2025agenticir}. For MRI, AgentMRI lets a vision-language model name the corruption and
dispatch a restoration model~\cite{saju2025agentmri}; MRI-AgentNet coordinates several such agents for motion,
undersampling and noise~\cite{saju2026mriagentnet}; ResMAP restores MRI with mixed artifacts by cascading prompt
retrieval~\cite{tang2025resmap}. Yet AgentMRI and MRI-AgentNet only look, through a vision-language model, and such models are least reliable on mild degradations~\cite{liu2025medqbench}.
In MRI quality assessment, MMRQA grades a scan better when its vision-language model is informed by the scan's global quality metrics~\cite{jia2025mmrqa}.

\begin{figure*}[t]
\centering
\includegraphics[width=\textwidth]{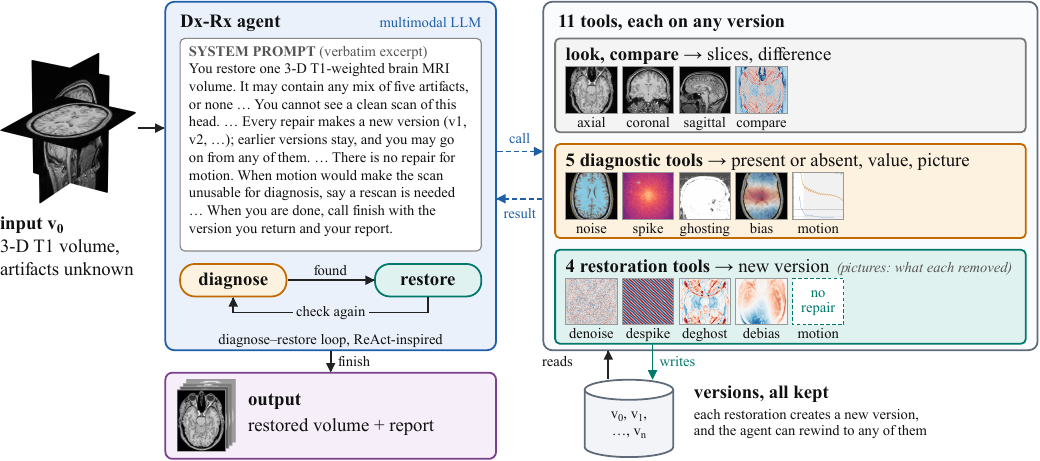}
\caption{Dx-Rx. The MLLM receives one 3-D volume with unknown artifacts and a system prompt (excerpt, verbatim) and
works in a diagnose--restore loop inspired by ReAct~\cite{yao2023react}: it calls tools on any version and reads their
results (text and pictures) before deciding again.}
\label{fig:dxrx}
\end{figure*}

We test the idea, \emph{measure before you repair}, on whole 3-D volumes with up to five artifacts. Our
main contributions are:
\begin{itemize}
\item Dx-Rx, an MLLM agent that measures with diagnostic tools, repairs, checks each repair, and can return to any
earlier version and change the order (Fig.~\ref{fig:dxrx}).
\item To our knowledge, the first agent to restore whole 3-D MRI volumes. Earlier MRI agents restore 2-D
slices~\cite{saju2025agentmri,saju2026mriagentnet}.
\item A radiology-style report that gives the measured evidence for each finding and explains each repair and its
order, and is checked deterministically against the run log for hallucinations.
\item An ablation showing how much the diagnostic tools, the loop and a larger MLLM each add to the agent.
\end{itemize}

\section{Method}
\label{sec:method}

Dx-Rx receives one 3-D T1-weighted volume ($192\times240\times240$, 1\,mm) that may contain any mix of five
artifacts, or none, and nothing else: no clean scan, no artifact labels and no file name. Its system prompt
(Fig.~\ref{fig:dxrx}) states the task, the scanner protocol (phase encoding along anterior--posterior, A--P), and that motion cannot be repaired: a rescan is
advised when the motion tool returns \calC{} or more (an error of \calCpct\,\% of white matter), the level at which a motion scan in the
Movement-Related Artefacts (MR-ART) dataset~\cite{narai2022} was as likely to be judged unusable by radiologists as usable. It gives no order of repairs and no hints about the data.

\begin{figure}[t]
\centering
\includegraphics[width=\columnwidth]{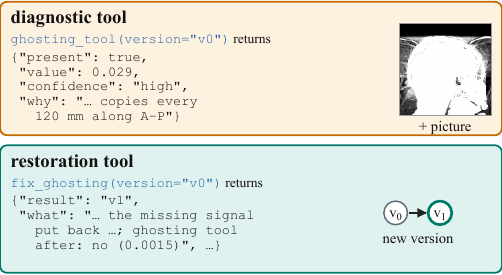}
\caption{What the agent receives from a tool (t0100, GPT-6.1 Terra, shortened). A diagnostic tool returns whether
its artifact is present, a value, a confidence, a reason and a picture of its evidence. A restoration tool returns
the new version and its diagnostic tool's new answer.}
\label{fig:tools}
\end{figure}

The MLLM acts only through eleven tools and \emph{finish} (Fig.~\ref{fig:dxrx}), each taking a version handle ($v_0$
is the input). \emph{Look} returns slices from 3-D volumes and the MLLM sees 2-D views, while tools work on whole 3-D volumes. \emph{Compare} returns the difference of
two versions and its root mean square (RMS). Each of the five diagnostic tools says whether one artifact is present or absent, with a
value, a confidence, a reason and a picture of its evidence. The noise tool estimates the standard deviation ($\sigma$) in flat tissue; the spike tool
finds isolated bright points in k-space~\cite{kao2000}; the ghosting tool finds faint copies of the head along A--P;
the bias tool fits an N4 field~\cite{tustison2010} over the brain, found with HD-BET~\cite{isensee2019}; the motion tool
measures head signal leaking into the air along A--P~\cite{zaitsev2015}. The four restoration tools remove spikes in
k-space; put back the signal that ghosting moved along A--P; denoise with DRUNet~\cite{zhang2022drunet}, a Gaussian denoiser, at the
measured $\sigma$ slice by slice along each axis, averaging the three resulting volumes (Rician noise is near-Gaussian in tissue~\cite{gudbjartsson1995}); and divide out the N4 field. Each repair runs its diagnostic tool before and after, so the agent sees at once
whether the artifact is gone. Fig.~\ref{fig:tools} shows what the agent receives. Nothing is trained. The tools are classical methods or published weights, each with one
threshold and one scale set on development volumes. \emph{Finish} ends the run.

Dx-Rx runs a diagnose--restore loop inspired by ReAct~\cite{yao2023react} (Fig.~\ref{fig:dxrx}): the MLLM reasons,
calls tools, reads their results (text and pictures) and decides again. Every version is kept and any tool can act on
any version, so the agent can check a repair by measuring it again, compare two versions, and return to an earlier
version to try another order, as AgenticIR reschedules after reflection~\cite{zhu2025agenticir}.

\emph{Finish} returns a version and a report under radiology headings, from examination to
limits~\cite{acr2025communication}. Each artifact gets a finding that starts with present, absent or uncertain and gives
the evidence. The repairs are listed in order, with their reasons. A model's explanation can differ from what it
did~\cite{turpin2023unfaithful}, and a clinical report must not state anything the measurements do not support. Every
report is therefore fact-checked deterministically against the run log: (a) the repairs it lists
are the returned version's history; (b) every number in its text was returned by a tool; (c) its written findings agree with its own yes/no answers. A report that passes all three is \emph{faithful}.

\section{Experiments}
\label{sec:experiments}

We used the public IXI T1-weighted scans (three hospitals, 1.5\,T and 3\,T)~\cite{ixi}. We computed each scan's brain mask with
HD-BET brain extraction~\cite{isensee2019}, removed the coil shading with N4~\cite{tustison2010}, resampled each volume to 1\,mm and
scaled it to white matter $=1$. Five artifacts were simulated on the whole volume as one 3-D acquisition, in physical
order: a bias field, rigid motion as k-space segments of moved copies~\cite{shaw2019}, ghosting and k-space
spikes~\cite{kao2000}, then Rician noise~\cite{gudbjartsson1995} and the magnitude. Ghosting and spikes use
TorchIO~\cite{perezgarcia2021}, with the image kept complex until the magnitude is taken.

Each artifact has three levels, mild, moderate and borderline, calibrated to real scans. For motion we used
\mrartSubj{} MR-ART subjects~\cite{narai2022}, each scanned once still and twice while moving, and radiologists
graded their \mrartScans{} motion scans good (\mrartGood), medium (\mrartMed) or bad, i.e.\ unusable (\mrartBad). For each
motion scan, we measured how much its image entropy~\cite{atkinson1997} rose over the same subject's still scan. We then
applied our motion to the same still scans at many strengths. The entropy rise grew linearly ($r=\calR$) with the
damage we applied, defined as the RMS change over the brain as a fraction of white-matter intensity. From this linear
fit, each real scan gets the damage at which our motion raises entropy by the same amount. On this damage scale, a
quarter of the medium scans lay below \calA, half below \calB{} and three quarters below \calC, and half of the bad
scans lay below \calD. These four values bound the three levels: mild (\calA--\calB), moderate (\calB--\calC) and
borderline (\calC--\calD). At
\calC, a real scan was as likely to be graded unusable as usable, so we set the agent's rescan line there. Ghosting and spikes, which have no
clinical grade, use motion's damage scale; noise follows clinical noise grades (total $\sigma$ 3.0--6.1\,\%
of white matter)~\cite{loizillon2024melba}; the bias stays within the fields N4 removed from IXI scans.

Four disjoint sets of IXI subjects were used. \nCal{} calibrated the levels. \nSweep{} more, each with single
artifacts, set each tool's threshold and scale. \nDev{} development volumes with mixed artifacts were used to develop
the agent. The test set has \nTest{} subjects, one volume each: \siteGuys{} from Guy's Hospital, \siteHH{} from
Hammersmith Hospital and \siteIOP{} from the Institute of Psychiatry. Five volumes have no artifact, 25 have one, 50 have two, 50 have three, 15 have four and 3 have all five. Each method ran
on it once after the agent was frozen. The files are blind: the agent cannot learn anything from a file name or
header since names are opaque and all volumes share one header, and no method sees the labels or the clean volumes.

The share of error removed is $1-\lVert y-r\rVert/\lVert x-r\rVert$ inside the brain (HD-BET mask), for input $x$,
output $y$ and the clean volume $r$. No tool repairs motion, so $r$ keeps the volume's own simulated motion. The
metric scores only what repairs can fix. It is defined for the \nRep{} test volumes with a repairable artifact. For the other 10 (5 clean and 5 motion-only), the input already equals $r$, so there is no error to remove. We report means $\pm$ standard deviation (std). We also count volumes made worse, correct diagnoses (the yes/no for 5 artifacts $\times$
\nTest{} volumes), faithful reports, cost, PSNR and SSIM~\cite{wang2004} inside the brain.
Methods are compared per volume with the Wilcoxon signed-rank
test~\cite{wilcoxon1945}, zero differences dropped, and Holm's adjustment~\cite{holm1979} across the models in each comparison.

Four questions organize the comparisons; Table~\ref{tab:methods} shows what each method can do.
\textbf{Q1: is an agent the smart choice?} Apply-all runs the four repairs on every volume in the fixed order: spike,
ghosting, noise, bias. The rule-based pipeline uses the same order but runs a repair only when its diagnostic tool says
yes. \textbf{Q2: how much can the restoration tools recover?} The oracle pipeline runs the same repairs in the same
order, but on the ground-truth artifacts instead of the diagnostic tools' results. With the diagnosis perfect, it
shows how much our restoration tools can recover. \textbf{Q3: which component of the agent matters?} In an ablation, three
variants of Dx-Rx remove (i) both the loop and the diagnostic tools, (ii) only the loop, or (iii) only the diagnostic tools. \emph{Single pass, no diagnostic tools}, as in AgentMRI~\cite{saju2025agentmri}: the MLLM sees the three planes once and names the repairs,
which run in the fixed order. \emph{Single pass + diagnostic tools:} the previous variant plus the five diagnostic tools'
results on the input. \emph{Loop, no diagnostic tools:} the Dx-Rx loop, which can look, compare and repair, but without its diagnostic tools' answers.
\textbf{Q4: does a larger model help?} We ran Dx-Rx with a smaller and a larger model of two families, Claude Haiku
5.5 and Sonnet 5.5, and GPT-6 Luna and GPT-6.1 Terra. We also ran the open-weights Qwen 3.8-27B.

\begin{table}[t]
\centering
\caption{Compared methods and their components.}
\label{tab:methods}
\small
\setlength{\tabcolsep}{3pt}
\begin{tabular}{@{}>{\raggedright\arraybackslash}p{30mm}ccccc@{}}
\toprule
Method & MLLM & Image & \shortstack{Diag.\\tools} & \shortstack{Checks,\\undoes} & \shortstack{Chooses\\order} \\
\midrule
Apply-all pipeline & -- & \no & \no & \no & \no \\
Rule-based pipeline & -- & \no & \yes & \no & \no \\
Oracle pipeline & -- & \no & truth & \no & \no \\
Single pass, no diag.\ tools & 1 call & \yes & \no & \no & \no \\
Single pass + diag.\ tools & 1 call & \yes & \yes & \no & \no \\
Loop, no diag.\ tools & loop & \yes & \no & \yes & \yes \\
\rowcolor{ours} \textbf{Dx-Rx (ours)} & loop & \yes & \yes & \yes & \yes \\
\bottomrule
\end{tabular}
\end{table}

\begin{figure*}[t]
\centering
\includegraphics[width=\textwidth]{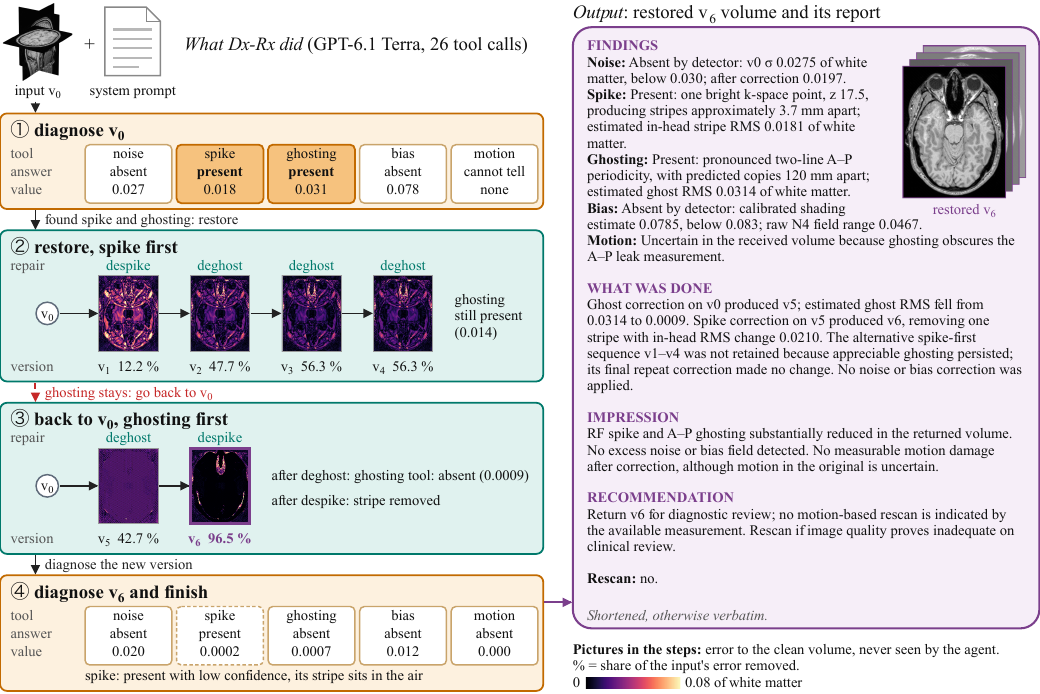}
\caption{One test volume (t0122: spike and ghosting) from input to output, GPT-6.1 Terra}
\label{fig:case}
\end{figure*}

\section{Results}
\label{sec:results}

\begin{table}[t]
\centering
\caption{Q1 and Q2: Pipelines and the agent. Error removed (\%) is the mean $\pm$ std over the \nRep{} volumes with a repairable
artifact. Made worse is out of \nTest{} volumes. PSNR (dB) and SSIM are means $\pm$ std inside the brain over the \nArt{} volumes
with an artifact. The oracle pipeline is told the true artifacts.}
\label{tab:q1}
\small
\setlength{\tabcolsep}{3.8pt}
\begin{tabular}{@{}l@{\hspace{3pt}}rcrr@{}}  
\toprule
Method & \multicolumn{1}{c}{\shortstack{Error\\removed (\%)}} & \multicolumn{1}{c}{\shortstack{Made\\\vphantom{d}worse}} & \multicolumn{1}{c}{PSNR} & \multicolumn{1}{c@{}}{SSIM} \\
\midrule
\rule{0pt}{\dimexpr\ht\strutbox+2pt\relax}No restoration\rule[-\dimexpr\dp\strutbox+2pt\relax]{0pt}{0pt} & 0 & -- & 32.3 $\pm$ 3.7 & 0.951 $\pm$ 0.036 \\
{\renewcommand{\arraystretch}{0.85}\begin{tabular}[c]{@{}l@{}}\rule{0pt}{\dimexpr0.85\ht\strutbox+2pt\relax}Apply-all\\pipeline\rule[-\dimexpr0.85\dp\strutbox+2pt\relax]{0pt}{0pt}\end{tabular}} & 52.1 $\pm$ 15.8 & 11 & 38.0 $\pm$ 3.4 & 0.981 $\pm$ 0.010 \\
{\renewcommand{\arraystretch}{0.85}\begin{tabular}[c]{@{}l@{}}\rule{0pt}{\dimexpr0.85\ht\strutbox+2pt\relax}Rule-based\\pipeline\rule[-\dimexpr0.85\dp\strutbox+2pt\relax]{0pt}{0pt}\end{tabular}} & 62.3 $\pm$ 17.9 & 0 & 40.1 $\pm$ 8.2 & 0.985 $\pm$ 0.012 \\
\rowcolor{ours} {\renewcommand{\arraystretch}{0.85}\begin{tabular}[c]{@{}l@{}}\rule{0pt}{\dimexpr0.85\ht\strutbox+2pt\relax}Dx-Rx,\\GPT-6.1 Terra\rule[-\dimexpr0.85\dp\strutbox+2pt\relax]{0pt}{0pt}\end{tabular}} & 64.5 $\pm$ 18.9 & 0 & 40.7 $\pm$ 9.1 & 0.985 $\pm$ 0.012 \\
\midrule
{\renewcommand{\arraystretch}{0.85}\begin{tabular}[c]{@{}l@{}}\rule{0pt}{\dimexpr0.85\ht\strutbox+2pt\relax}\textit{Oracle pipeline}\\\textit{(true artifacts)}\rule[-\dimexpr0.85\dp\strutbox+2pt\relax]{0pt}{0pt}\end{tabular}} & \textit{64.6 $\pm$ 15.6} & \textit{0} & \textit{40.4 $\pm$ 8.2} & \textit{0.985 $\pm$ 0.012} \\
\bottomrule
\end{tabular}
\end{table}

\begin{table}[t]
\centering
\caption{Q3: What each component adds. Each value is the error removed (\%), as in Table~\ref{tab:q1}.}
\label{tab:q3}
\small
\setlength{\tabcolsep}{11pt}
\begin{tabular}{@{}lrrr@{}}
\toprule
Variant & \multicolumn{1}{c}{\shortstack{Claude\\Haiku 5.5}} & \multicolumn{1}{c}{\shortstack{GPT-6\\Luna}} & \multicolumn{1}{c@{}}{\shortstack{\raisebox{0pt}[\height][0pt]{Qwen}\\3.8-27B}} \\
\midrule
{\renewcommand{\arraystretch}{0.85}\begin{tabular}[c]{@{}l@{}}\rule{0pt}{\dimexpr0.85\ht\strutbox+2pt\relax}Single pass,\\no diag.\ tools\rule[-\dimexpr0.85\dp\strutbox+2pt\relax]{0pt}{0pt}\end{tabular}} & 19.7 $\pm$ 25.8 & 19.6 $\pm$ 24.4 & 3.3 $\pm$ 13.0 \\
{\renewcommand{\arraystretch}{0.85}\begin{tabular}[c]{@{}l@{}}\rule{0pt}{\dimexpr0.85\ht\strutbox+2pt\relax}Single pass\\+ diag.\ tools\rule[-\dimexpr0.85\dp\strutbox+2pt\relax]{0pt}{0pt}\end{tabular}} & 61.7 $\pm$ 18.6 & 60.0 $\pm$ 21.3 & 61.7 $\pm$ 18.6 \\
{\renewcommand{\arraystretch}{0.85}\begin{tabular}[c]{@{}l@{}}\rule{0pt}{\dimexpr0.85\ht\strutbox+2pt\relax}Loop,\\no diag.\ tools\rule[-\dimexpr0.85\dp\strutbox+2pt\relax]{0pt}{0pt}\end{tabular}} & 60.1 $\pm$ 21.1 & 62.0 $\pm$ 18.9 & 59.3 $\pm$ 22.0 \\
\rowcolor{ours} {\renewcommand{\arraystretch}{0.85}\begin{tabular}[c]{@{}l@{}}\rule{0pt}{\dimexpr0.85\ht\strutbox+2pt\relax}Dx-Rx (loop\\+ diag.\ tools)\rule[-\dimexpr0.85\dp\strutbox+2pt\relax]{0pt}{0pt}\end{tabular}} & 62.2 $\pm$ 18.0 & 62.4 $\pm$ 17.8 & 63.1 $\pm$ 18.3 \\
\bottomrule
\end{tabular}
\end{table}

\begin{table}[t]
\centering
\caption{Q4: Dx-Rx with smaller and larger MLLMs. Error removed is as in Table~\ref{tab:q1}. Vs.\
rule-based counts volumes with more / less error removed than the rule-based pipeline ($^*$significant after
Holm's adjustment). Diag.\ and Faithful are the shares of correct diagnoses and of faithful reports.
Cost is the mean per volume.}
\label{tab:q4}
\small
\setlength{\tabcolsep}{6.1pt}
\begin{tabular}{@{}lrrrrr@{}}
\toprule
Model & \multicolumn{1}{c}{\shortstack{Error\\removed (\%)}} & \multicolumn{1}{c}{\shortstack{vs.\ rule-\\based}} & \multicolumn{1}{c}{\shortstack{\raisebox{0pt}[\height][0pt]{Diag.}\\(\%)}} & \multicolumn{1}{c}{\shortstack{Faithful\\(\%)}} & \multicolumn{1}{c@{}}{\shortstack{Cost\\(\$)}} \\
\midrule
{\renewcommand{\arraystretch}{0.85}\begin{tabular}[c]{@{}l@{}}\rule{0pt}{\dimexpr0.85\ht\strutbox+2pt\relax}Claude\\Haiku 5.5\rule[-\dimexpr0.85\dp\strutbox+2pt\relax]{0pt}{0pt}\end{tabular}} & 62.2 $\pm$ 18.0 & 11 / 14 & 90.9 & 95.9 & 0.003 \\
{\renewcommand{\arraystretch}{0.85}\begin{tabular}[c]{@{}l@{}}\rule{0pt}{\dimexpr0.85\ht\strutbox+2pt\relax}GPT-6\\Luna\rule[-\dimexpr0.85\dp\strutbox+2pt\relax]{0pt}{0pt}\end{tabular}} & 62.4 $\pm$ 17.8 & 8 / 9 & 91.5 & 96.6 & \textbf{0.002} \\
{\renewcommand{\arraystretch}{0.85}\begin{tabular}[c]{@{}l@{}}\rule{0pt}{\dimexpr0.85\ht\strutbox+2pt\relax}Qwen\\3.8-27B\rule[-\dimexpr0.85\dp\strutbox+2pt\relax]{0pt}{0pt}\end{tabular}} & 63.1 $\pm$ 18.3 & 13 / 6 & 91.2 & 87.2 & 0.010 \\
\midrule
{\renewcommand{\arraystretch}{0.85}\begin{tabular}[c]{@{}l@{}}\rule{0pt}{\dimexpr0.85\ht\strutbox+2pt\relax}GPT-6.1\\Terra\rule[-\dimexpr0.85\dp\strutbox+2pt\relax]{0pt}{0pt}\end{tabular}} & \textbf{64.5 $\pm$ 18.9} & \textbf{8 / 0}$^*$ & \textbf{92.0} & 97.3 & 0.017 \\
{\renewcommand{\arraystretch}{0.85}\begin{tabular}[c]{@{}l@{}}\rule{0pt}{\dimexpr0.85\ht\strutbox+2pt\relax}Claude\\Sonnet 5.5\rule[-\dimexpr0.85\dp\strutbox+2pt\relax]{0pt}{0pt}\end{tabular}} & 62.2 $\pm$ 18.8 & 13 / 13 & 89.7 & \textbf{99.3} & 0.035 \\
\bottomrule
\end{tabular}
\end{table}

\textbf{Q1: an agent can replace hand-set rules.} Running every repair made \harmApplyAll{} volumes worse
(Table~\ref{tab:q1}). Gating each repair by its diagnostic tool removed this harm and raised the mean from
\remApplyAll{} to \remRule. Dx-Rx needed no hand-set order or gating rule beyond the tools, and with none of the five MLLMs did it make a
volume worse. With every MLLM, it gave the same result as the rule-based pipeline on most volumes
(at least \sameLo{} of \nRep).

\textbf{Q2: the tools set the ceiling.} Given the true artifacts, the same tools in the fixed order removed
\remOracle. Without being told, Dx-Rx nearly matched it, removing \remTerra{} with GPT-6.1 Terra, and with no MLLM did it differ
significantly from the oracle ($p\ge\oracleMinP$). Even with a perfect diagnosis, the repairs leave the rest of the error.

\textbf{Q3: measuring is what makes it work.} Looking at the image alone, a single-pass MLLM removed at most
\remImgHi{} of the error (Table~\ref{tab:q3}). With each MLLM, Dx-Rx was better than this single pass on at least \imgHaikuB{} of the
\nRep{} volumes and worse on none. Given the diagnostic tools' results, the same single pass also came close but stayed
below Dx-Rx with every MLLM, though not significantly ($p\ge\oneMinP$). Without the diagnostic tools, the loop also stayed below Dx-Rx with every MLLM.
The difference was significant with Claude Haiku 5.5 and Qwen 3.8-27B, but not with GPT-6 Luna. The diagnostic tools carry most of the gain in error removed. The loop adds feedback:
it sees each repair's result, so it can check it, go back and change the order. For example, the fixed order repairs the spike before
ghosting. In \ghostN{} test runs, Dx-Rx repaired ghosting first instead and these runs beat the oracle
\ghostBeat{} times and were never worse.

Fig.~\ref{fig:case} shows one such run. The agent first repaired the spike and then ghosting like the fixed order,
but ghosting remained even after two more tries. It went back to the input and repaired ghosting first and then the
spike, removing \caseDxPct\,\% of the error against \caseFixedPct\,\% for the fixed order, and its report says why
it changed course.

\textbf{Q4: with a larger model, Dx-Rx nearly reaches the oracle.} With GPT-6.1 Terra it removed \remTerra,
nearly matching the \remOracle{} of the oracle pipeline, and significantly more than the smaller \mbox{GPT-6} Luna of
the same family, which removed \remLuna{} ($p=\famTerraP$). It was also the only MLLM never worse than the rule-based pipeline on any volume,
and it was better on \terraBetter{} volumes (Table~\ref{tab:q4}, $p=\terraP$).

In the Claude 5.5 family, the larger Sonnet removed only about as much as Haiku (both \remSonnet,
$p=\famSonnetP$), but it wrote the most faithful reports of all five MLLMs
(\faithHi{} of \nTest{} reports).

\section{Conclusion}
\label{sec:conclusion}
Dx-Rx restores whole 3-D brain MRI with several artifacts as well as or better than an expert-designed rule-based pipeline built from
the same tools, without hand-set rules for which repair to run or in what order, and also explains each decision in a
radiology-style report. Its strength comes from measuring before repairing. An MLLM that only
looks at slices removes little, while the same MLLM with diagnostic tools comes close to the expert pipeline, and the loop lets it
notice when a repair did not work and change course. With a larger MLLM, Dx-Rx nearly reached the oracle pipeline that is told the ground truth
artifacts. The tools set the ceiling. It also worked well with an open-weight model, which a hospital can run on site to keep scans in-house and comply with patient privacy regulations. 

\section{Compliance with Ethical Standards}
\label{sec:ethics}
This research study was conducted retrospectively using human subject data made available in open access by the IXI
dataset~\cite{ixi} (CC BY-SA 3.0) and the MR-ART dataset (OpenNeuro ds004173, CC0)~\cite{narai2022}. Ethical approval was
not required as confirmed by the licenses attached with the open access data.

\section{Acknowledgments}
\label{sec:ack}
We thank the IXI project and the MR-ART authors, through OpenNeuro, for making their data available. The authors declare no conflicts of interest.

\bibliographystyle{IEEEbib}
\bibliography{refs}

\end{document}

%% file: numbers.tex
\newcommand{\nTest}{148}
\newcommand{\nRep}{138}
\newcommand{\nArt}{143}

\newcommand{\remApplyAll}{52.1\,\%}
\newcommand{\remRule}{62.3\,\%}
\newcommand{\remOracle}{64.6\,\%}
\newcommand{\remTerra}{64.5\,\%}
\newcommand{\harmApplyAll}{11}

\newcommand{\remImgHi}{19.7\,\%}

\newcommand{\terraBetter}{8}

\newcommand{\terraP}{0.04}

\newcommand{\sameLo}{112}

\newcommand{\oracleMinP}{0.06}

\newcommand{\imgHaikuB}{114}

\newcommand{\oneMinP}{0.07}

\newcommand{\faithHi}{147}

\newcommand{\ghostN}{18}
\newcommand{\ghostBeat}{16}

\newcommand{\caseDxPct}{96.5}
\newcommand{\caseFixedPct}{47.7}
\newcommand{\calA}{0.019}
\newcommand{\calB}{0.028}
\newcommand{\calC}{0.033}
\newcommand{\calD}{0.055}
\newcommand{\calCpct}{3.3}
\newcommand{\calR}{0.92}
\newcommand{\mrartSubj}{30}
\newcommand{\mrartScans}{60}
\newcommand{\mrartGood}{10}
\newcommand{\mrartMed}{17}
\newcommand{\mrartBad}{33}
\newcommand{\nDev}{60}
\newcommand{\nSweep}{10}
\newcommand{\nCal}{10}
\newcommand{\siteGuys}{82}
\newcommand{\siteHH}{47}
\newcommand{\siteIOP}{19}
\newcommand{\remSonnet}{62.2\,\%}

\newcommand{\famTerraP}{0.02}

\newcommand{\famSonnetP}{0.85}

\newcommand{\remLuna}{62.4\,\%}